\documentclass[reprint,showpacs,showkeys,preprintnumbers,amsmath,amssymb,aip,apl]{revtex4-1}

\usepackage{graphicx}
\usepackage{subfigure}
\usepackage{float}
\usepackage{color}
\usepackage{soul}
\usepackage{nicefrac}

\begin{document}


\title{Loss of anisotropy in strained ultrathin epitaxial L1$_0$ Mn-Ga films} 

\author{Albrecht~K\"{o}hler}
\email{albrecht.koehler@cpfs.mpg.de}
\affiliation{Max Planck Institute for Chemical Physics of Solids,
             01187 Dresden, Germany.}

\author{Ivan Knez}
\affiliation{IBM Almaden Research Center, San Jose, CA 95120, USA.}
                          
\author{Daniel~Ebke}
\affiliation{Max Planck Institute for Chemical Physics of Solids,
             01187 Dresden, Germany.}
                        
\author{Claudia~Felser}
\affiliation{Max Planck Institute for Chemical Physics of Solids,
             01187 Dresden, Germany.}
						
\author{Stuart~S.P.~Parkin}
\email{Stuart.Parkin@us.ibm.com}
\affiliation{IBM Almaden Research Center, San Jose, CA 95120, USA.}

\date{\today}

\begin{abstract}

In this work we are investigating the effect of strain in ultrathin Mn-Ga thin films on the magnetic properties at room temperature. Two different Mn-Ga compositions of which one is furthermore doped with Co were grown on Cr buffered MgO (001) substrates. Films with a thickness below 12nm are highly strained and it was observed that the ratio $c/a$ vs. thickness is depending on composition. Using $c/a$ as an order parameter, the PMA is shown to be drastically reduced with increasing strain. These findings should be considered when generalizing and downscaling results obtained from films $>$20nm. Furthermore it has been shown, that the strain can be reduced by introducing an additional Pt buffer and thus maintaining a high PMA for a thickness as low as 6nm.
\end{abstract}



\maketitle

\begin{figure}
\renewcommand{\baselinestretch}{1.0}
\includegraphics[width=8cm]{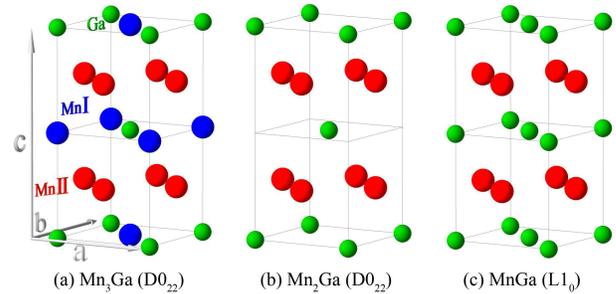}
\caption{Range of the binary Mn-Ga compound}
\label{unit_cell}
\end{figure}

Mn-Ga compounds are in the focus of research as they are a promising candidate for spintronic applications, such as spin-transfer-torque magnetic random access memory (STT-RAM)\cite{Slonczewski96,Felser11}. Mn$_3$Ga in the cubic phase is predicted to be a fully compensated nearly half-metallic ferrimagnet with a spin polarization of $88\%$\cite{Wurmehl06,Balke07,Winterlik08}. However, the compound is only stable in the tetragonal phase, which shows hard out-of-plane magnetic properties\cite{Kren70}. Furthermore the Curie Temperature is $T_C=730^\circ C$. The key for the applicability of Mn-Ga in these devices is the observed giant perpendicular magnetic anisotropy (PMA), which is necessary for long term thermal stability of magnetic tunnel junctions (MTJs). Non-volatility of the stored information requires that K$_U$V$\geq$40-100k$_B$T, where K$_U$ is the effective anisotropy and V is the unit cell volume.\\
The composition of Mn$_x$Ga$_{1-x}$ ranges from the instable cubic D0$_3$ phase (x$=$0.75), which is the binary equivalent of the cubic Heusler $L2_1$, to the tetragonal D0$_{22}$ phase (0.65$<$x$<$0.75) and lastly to the Mn poor L1$_0$ phase of MnGa. Through reduction of Mn below Mn$_2$Ga a stable tetragonal L1$_0$ phase is attained, which is practically a D0$_{22}$ with reduced symmetry due to the (partial) occupation of Ga atoms at the Wyckoff 2b positions\cite{Felser11}, thus halving the lattice constant $c$. All phases are shown in fig.\ref{unit_cell}. The blue (MnI) and red (MnII) positions of the Mn atoms are inequivalent and with the reduction of Mn from Mn$_3$Ga, mainly but not exclusively MnI is removed\cite{Winterlik08}. However, the bulk vacancy distribution model is not entirely consistent with the thin film model\cite{Mizukami12} and the detailed mechanism is still unclarified.
Many groups have been investigating varying composition ranges in bulk 
and thin films\cite{Mizukami12} ranging from Mn$_{1-x}$Ga$_x$ (x$=$0.6)\cite{Krishnan92} to Mn$_{3-x}$Ga (0$\leq$x$\leq$1)\cite{Mizukami09,Wu11,Coey11_02,Glas13,Glas13_02}. The reports about Mn-Ga however are mostly limited to a film thickness of 20-100nm and don't examine the effects of reducing the layer thickness. This is crucial for the implementation of this Heusler into devices with high information density capable of competing with existing technologies. To obtain a high magnetization (per volume) M with Mn-Ga a composition close to Mn$_2$Ga was targeted here, where the magnetic moment of MnII is not compensated due to a lack of MnI.

Offstoichiometrical Mn$_2$Ga films were prepared using the magnetron sputtering technique with an Ar pressure of 10$^{-3}$mbar. The base pressure was less than 10$^{-10}$mbar. The deposition steps on single crystalline (001)-MgO substrates were:
\begin{center}
\textbf{for Cr buffered films:}\\
\small{40nm Cr/Mn$_{1.9}$Co$_{0.1}$Ga/2h$@$400$^\circ$C/3nm Ta}\\
and\\
\small{40nm Cr/Mn$_{1.86}$Ga/2h$@$400$^\circ$C/3nm Ta}\\
\end{center}
\begin{center}
\textbf{for Cr/Pt buffered films:}\\
\small{10nm Cr/1h$@$700$^\circ$C/10nm Pt $@$500$^\circ$C/Mn$_{1.9}$Co$_{0.1}$Ga $@$450$^\circ$C/3nm Ta}
\end{center}
A step was conducted at room temperature when no explicit temperature is denoted. The Ta layer serves as protective capping. The composition of the films was analyzed by RBS/PIXE with a precision down to 1$\%$at. The lattice constant of MgO (a$=$4.21$\mathring{A}$) leads to an epitaxial 45$^\circ$ inplane rotated growth of Cr (a$=$2.88$\mathring{A}$). The Pt buffer (a$=$3.92$\mathring{A}$) then again grows 45$^\circ$ rotated on Cr.\\
When Mn$_{1.9}$Co$_{0.1}$Ga samples were deposited on the Pt buffer at room temperature they were nonmagnetic even after post annealing at 400$^\circ$C or 450$^\circ$C. Therefore those samples were deposited at 450$^\circ$C, which has been found to be the optimal substrate temperature\cite{Li13}.\\
Cu K$_{\alpha 1}$ radiation was used for the structural analysis of the samples. In accordance with the chosen D0$_{22}$ unit cell in fig.\ref{unit_cell} the (011) reflex was scanned to estimate the structure of the films. An absence of this reflex would suggest the L1$_0$ structure, because there the structure factor is zero. 
For the sake of consistency and to reduce confusion we throughoutly index the planes refering to the unit cell shown in fig.\ref{unit_cell}, e.g. the L1$_0$-(001) reflex is the same as D0$_{22}$-(002).\\
For the estimation of the inplane lattice constant $a$, an energy-dispersive 1-D detector was utilized to create a 2D $2\theta/\theta-\chi$ scan map in which the (112)-reflection appears as a 3D gaussian shape. The (112)-reflex was chosen, since its intensity is the strongest thus making the investigation by XRD down to a thickness of 2nm possible (to prevent confusion it has to be mentioned that some databases identify the (022) reflex as the one with the highest intensity, since they refer to the Heusler-type L2$_1$ unit cell). For the films on Cr/Pt-buffer the (224)-reflex had to be chosen for the estimation of $a$ due to peak overlap of the sample-(112) with the Pt-(111) reflex at 2$\theta$$=$39.8$^\circ$. Due to its reduced intensity it was not possible to appropriately estimate both lattice constants of the 3nm Cr/Pt-buffered Mn$_{1.9}$Co$_{0.1}$Ga film.

\begin{figure}[htb]
\centering
	 \includegraphics[width=2.77cm]{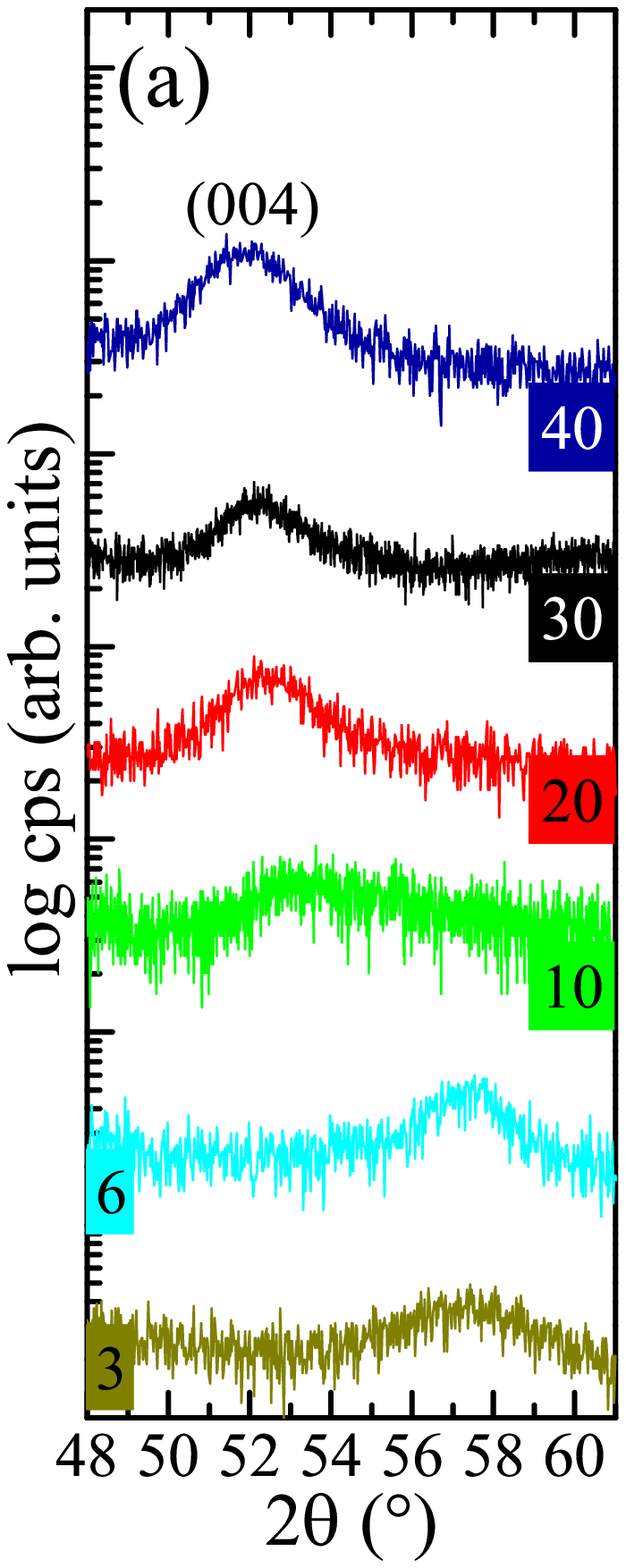}
		\hspace{0.05cm}
	 \includegraphics[width=2.3cm]{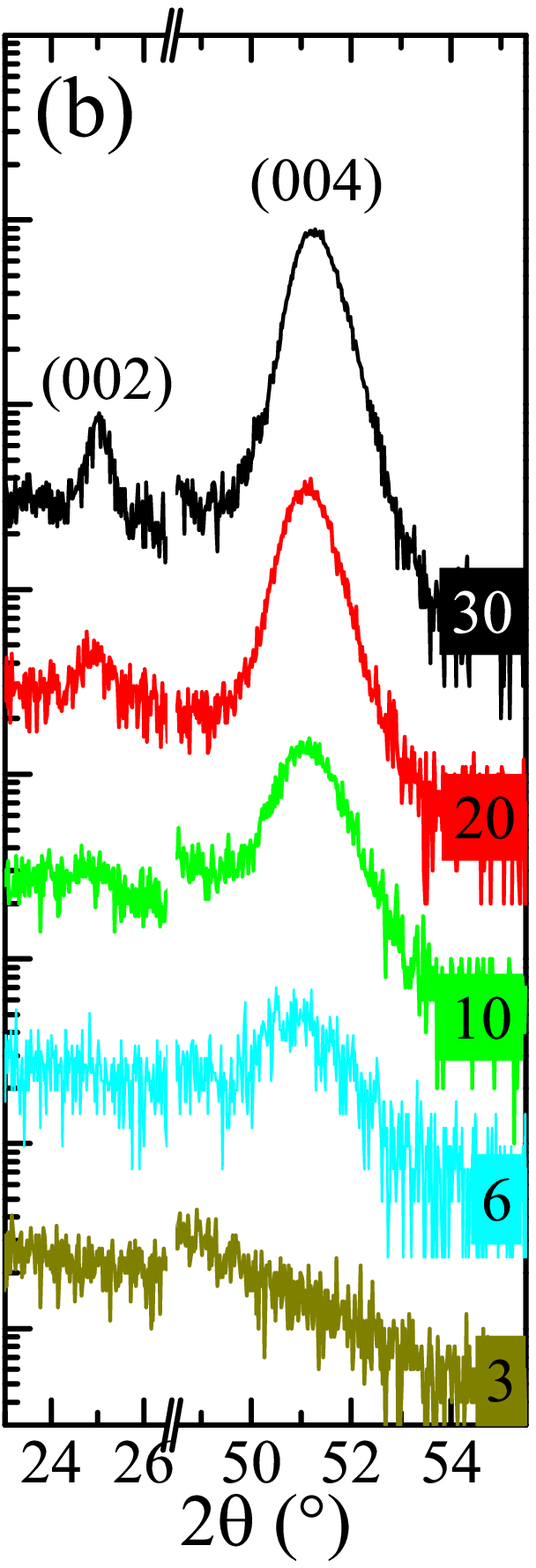}
		\hspace{0.05cm}
	 \includegraphics[width=3.04cm]{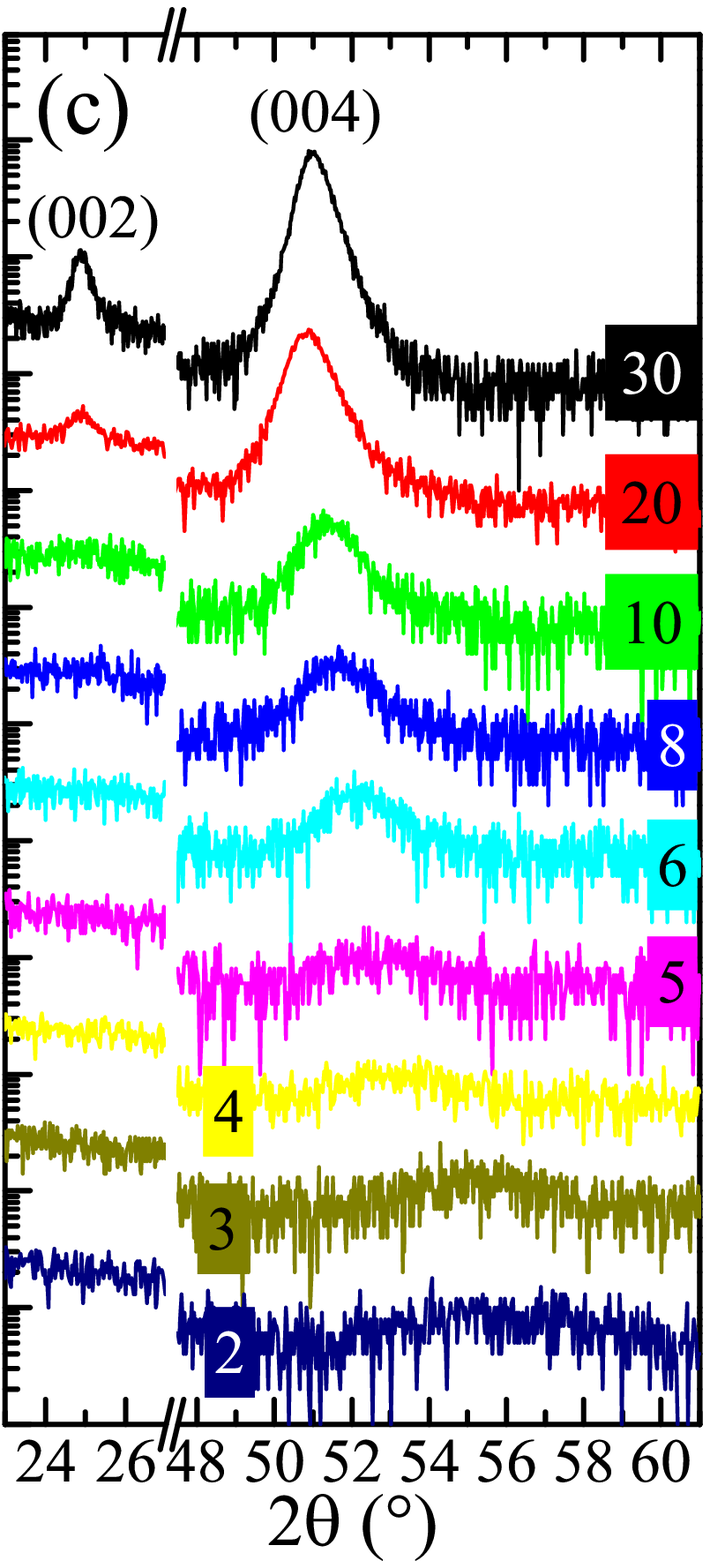}
   \caption{(color online), $2\theta/\theta$ scan excerpt of the (002)/(004) reflexes of (a) Mn$_{1.9}$Co$_{0.1}$Ga on Cr buffer, no (002) was observed; (b) Mn$_{1.9}$Co$_{0.1}$Ga on Cr/Pt buffer; (c) Mn$_{1.86}$Ga on Cr. The numbers denote the film thickness in nm.}
\label{XRD}
\end{figure}

\begin{figure}[htb]
\centering
   \includegraphics[width=4.25cm]{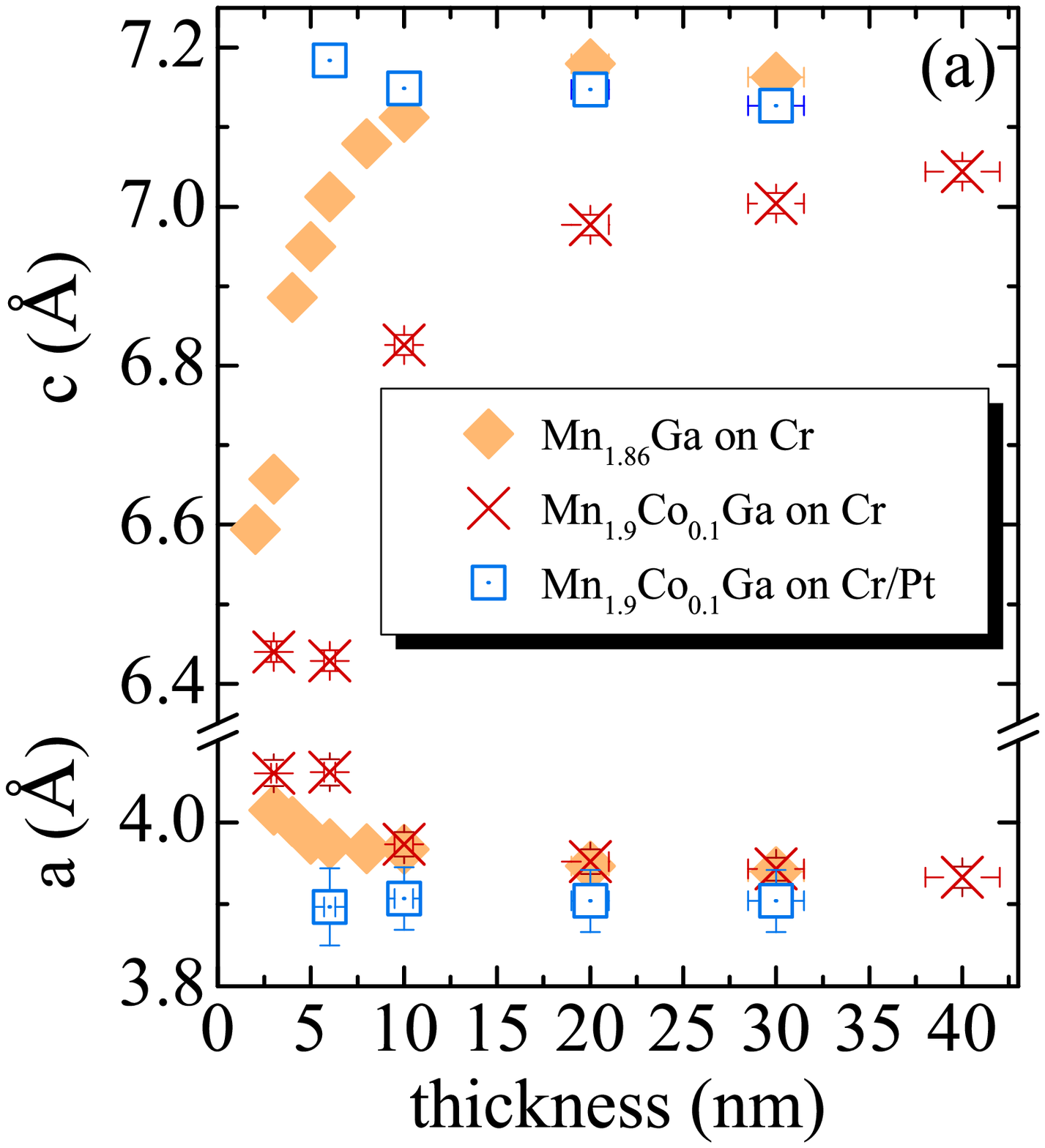}
	 \includegraphics[width=4.25cm]{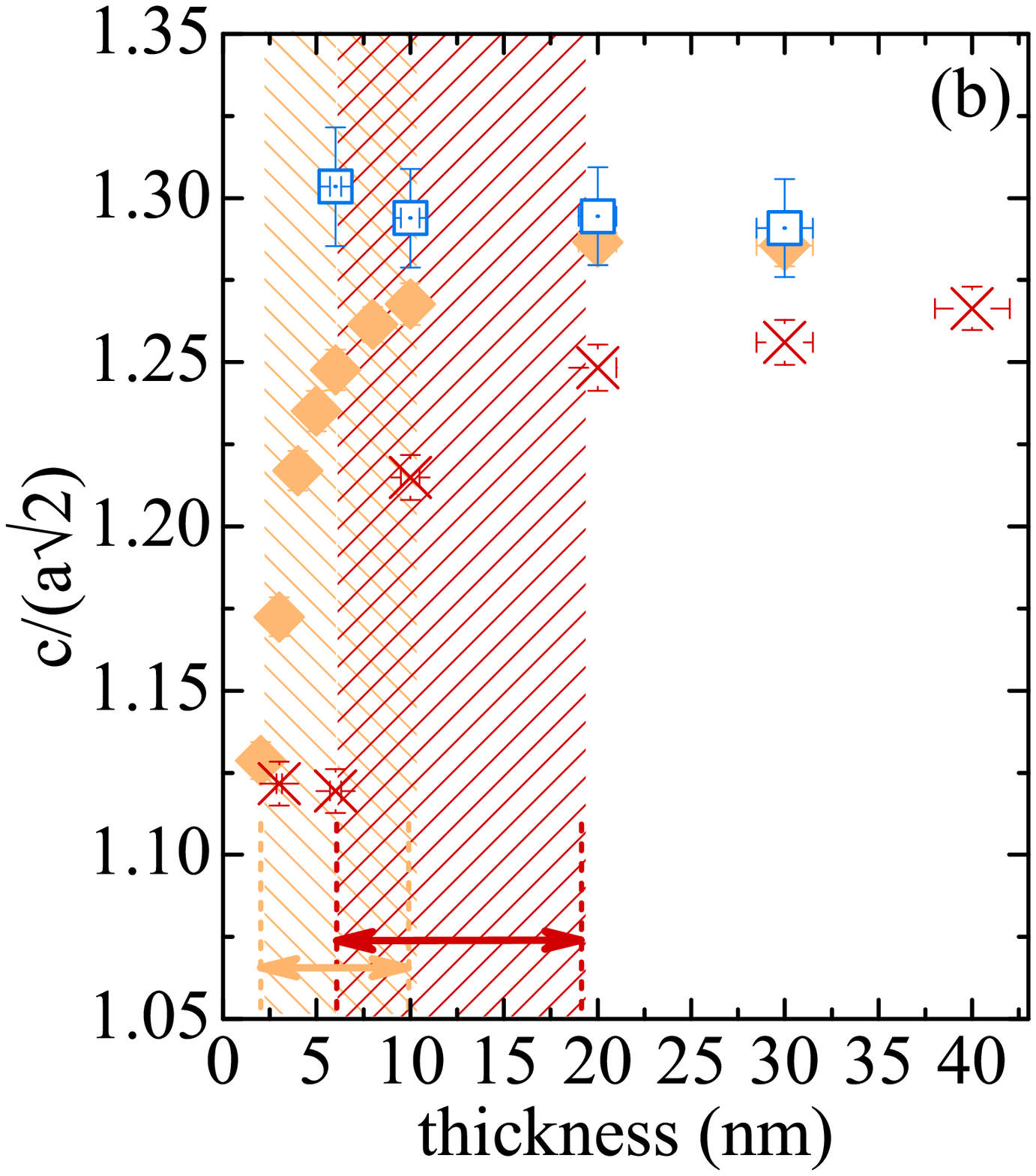}
   \caption{(color online) (a) lattice constants $c$ and $a$ versus thickness, arrow bars are included in all data points; (b) ratio $\nicefrac{c}{a \sqrt{2}}$ versus film thickness. The regions of a large strain gradient are denoted by shaded areas and arrows schematically. The Cr/Pt buffer films are unstrained within the whole range, whereas the Cr buffered compositions each show different strain vs. thickness behaviour.}
\label{fig:lattice}
\end{figure}

Fig.\ref{XRD} shows excerpts of the $2\theta/\theta$ scans of the films. In the full range only the (001)/(002) substrate and (002) Cr reflexes were visible apart from the (002) and (004) reflexes of the Mn-Ga(-Co) film. For the Cr/Pt buffered samples also the (002) Pt reflex appeared. All films t$\leq$10nm show no (002) superlattice peak, because the signal is too weak. An important observation is the complete absence of the (002) superlattice peak for Cr buffered Mn$_{1.9}$Co$_{0.1}$Ga, whereas they appear for the same composition on Cr/Pt buffer. This indicates a different ordering, that is probably related to the different deposition temperatures and not to the buffer layer.\\When changing the measurement geometry, none of the films showed the (011) reflex, so all films crystallized in the L1$_0$ structure. The lattice constants of the Cr buffer is $a_{Cr}=$2.885$\mathring{A}$ and for the Pt buffer $a_{Pt}=$3.946$\mathring{A}$, both in agreement with bulk values from literature, $a^{lit}_{Cr}=$2.88$\mathring{A}$, $a^{lit}_{Pt}=$3.92$\mathring{A}$. Therefore, all films show excellent epitaxial growth. A scan for the (111) Pt reflex proved its 45$^\circ$ rotated growth on the Cr buffer.\\
Fig.\ref{fig:lattice} shows the lattice constants and its ratios. It is noteworthy that the Cr/Pt-buffered thin films show a slight tetragonal elongation in $c$-direction when reducing the thickness. The lattice mismatch with Pt is $\Delta a$=$0.5\%$ for all films.\\
The Cr-buffered films are highly strained: When reducing the thickness of Mn$_{1.9}$Co$_{0.1}$Ga [Mn$_{1.86}$Ga] from 30nm down to 3nm, $c$ decreases by $8.5\%$ [$8.0\%$] while $a$ increases by $3.2\%$ [$4.6\%$] to values closer to the $45^\circ$ rotated Cr lattice constant $2.88\mathring{A}\cdot\sqrt{2}=4.08\mathring{A}$. The lattice mismatch of the 3nm Mn$_{1.9}$Co$_{0.1}$Ga [3nm Mn$_{1.86}$Ga] with the Cr buffer is $\Delta a=0.3\%$ [$\Delta a=1.2\%$]. Noteworthy are the different regions of large strain gradient: Mn$_{1.9}$Co$_{0.1}$Ga attains highest strain earlier when reducing thickness. We believe this is related to the tendency of Mn-Ga compounds to form a cubic phase when Co is added\cite{Ouardi12}. This is promoting the strain which is nothing else than the collapse of the tetragonal distortion towards a cubic phase.

\begin{figure}[htb]
\centering
   \includegraphics[width=8.5cm]{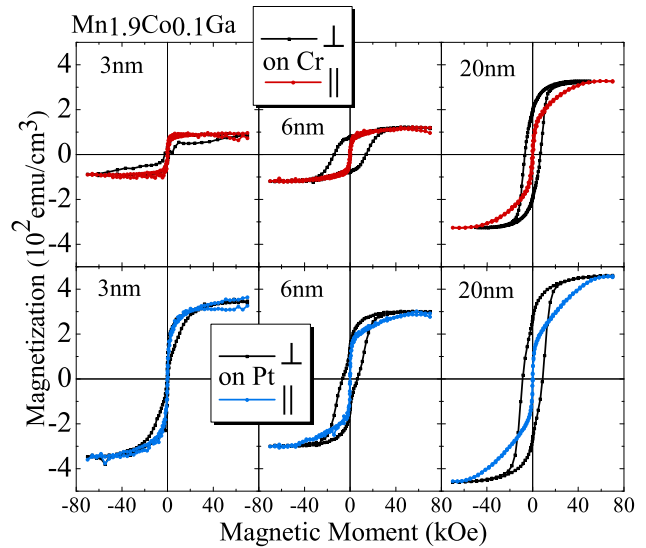}
   \caption{Hysteresis curves for Mn$_{1.9}$Co$_{0.1}$Ga for various thicknesses deposited on Cr (top) and Cr/Pt (bottom) buffer layers. The superimposed magnetic moment can clearly visible for the Cr/Pt buffered films.}
\label{fig:M_Cr1_Pt}
\end{figure}

\begin{figure}[htb]
\centering
   \includegraphics[width=8.5cm]{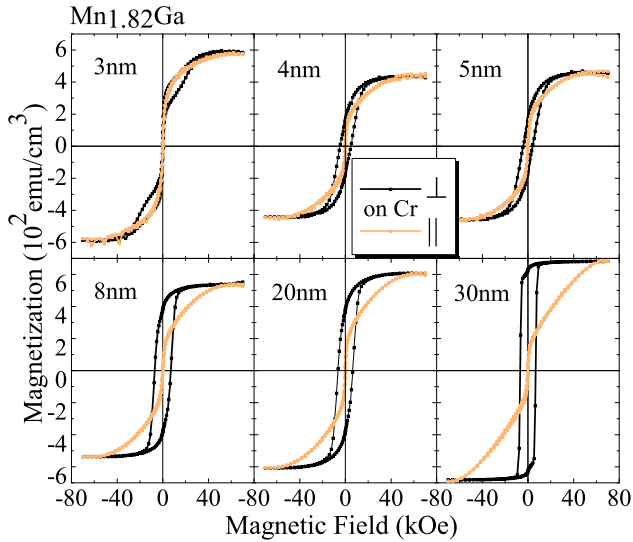}
   \caption{Hysteresis curves for Mn$_{1.86}$Ga for various thicknesses deposited on a Cr buffer layer.}
\label{fig:M_Cr2}
\end{figure}

\begin{figure}[htb]
\centering
	 \includegraphics[width=4.2cm]{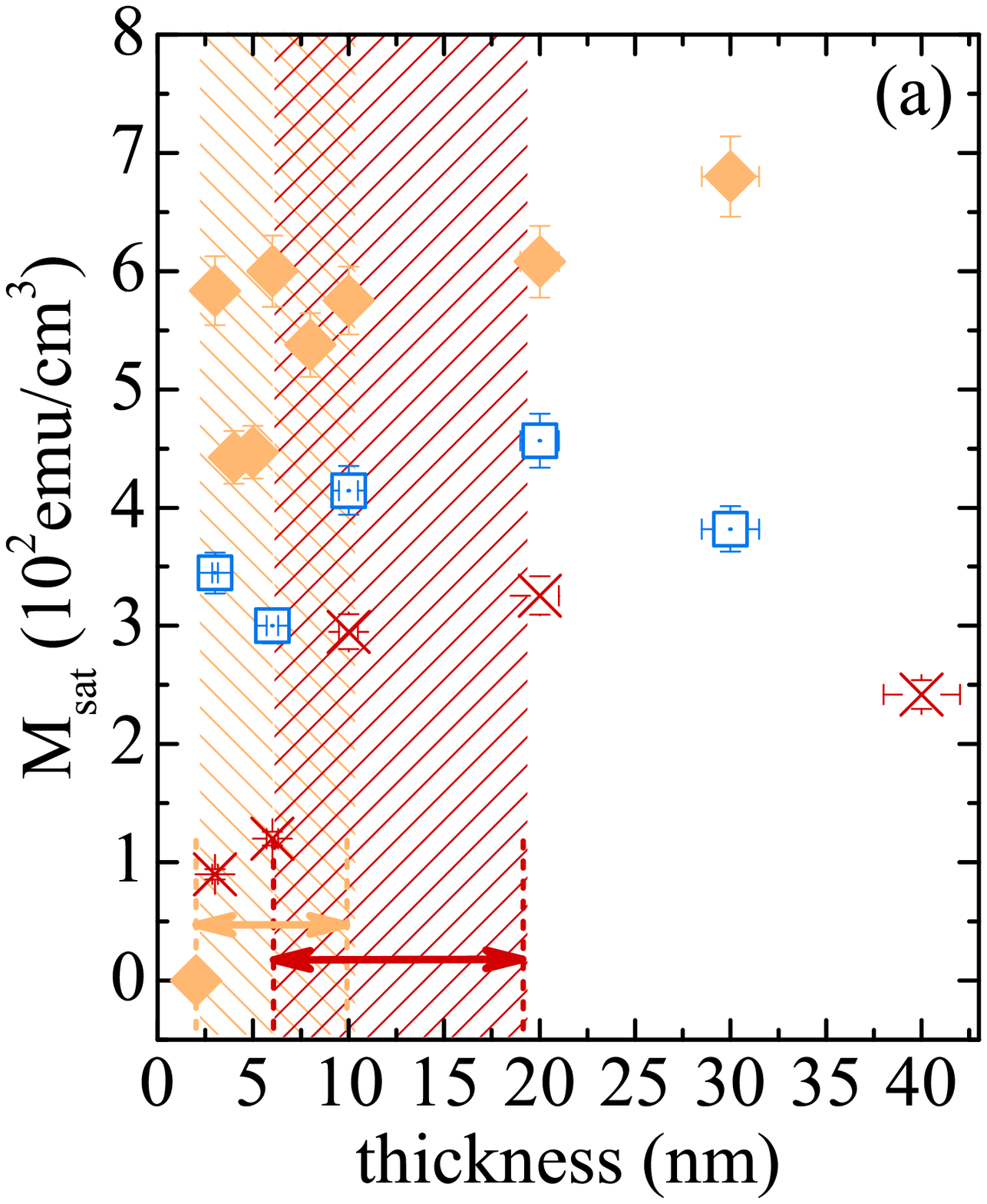}
   \includegraphics[width=4.3cm]{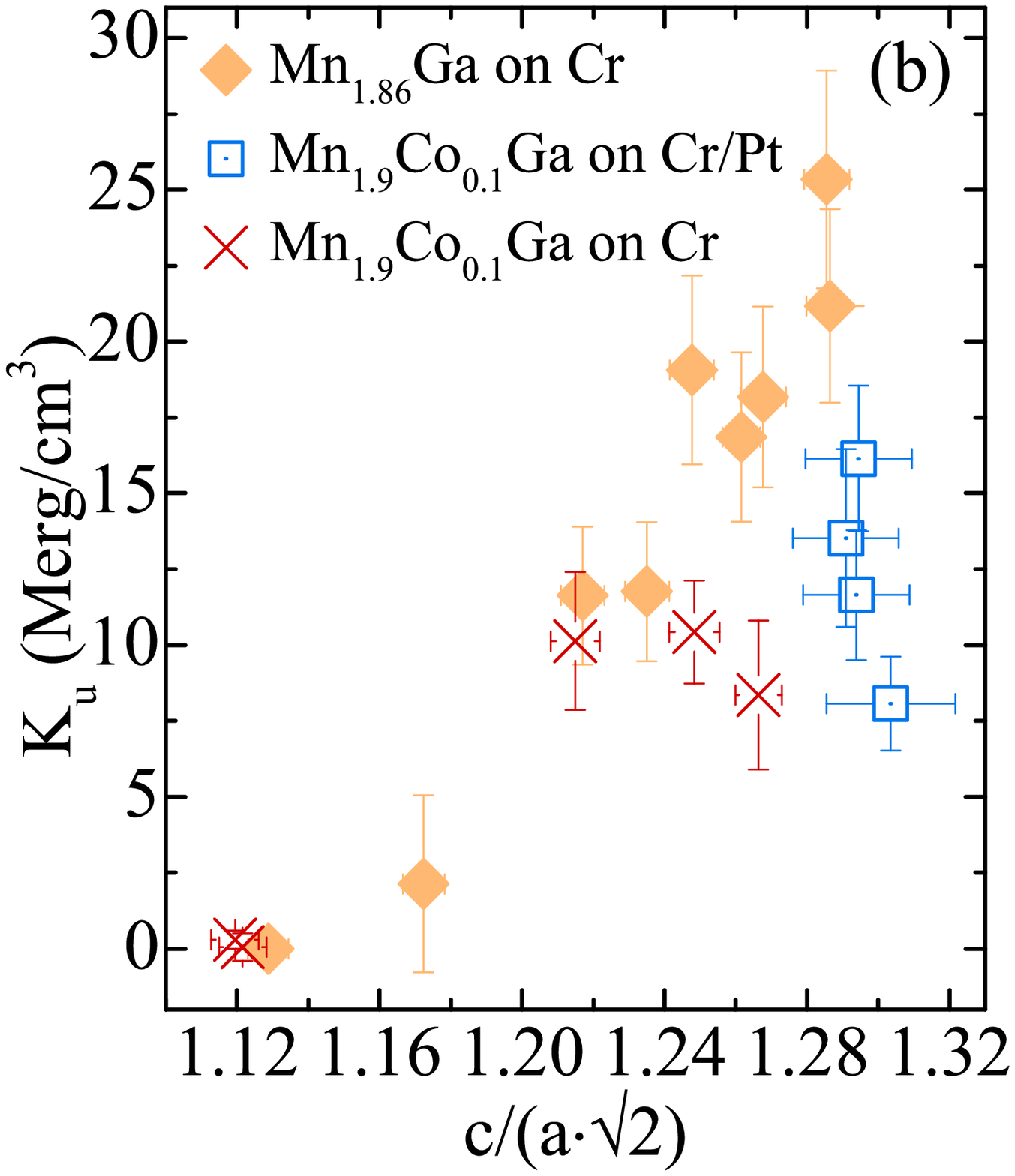}
   \caption{(a) M$_s$ per volume, the shaded regions are the same as in fig.\ref{fig:lattice}; (b) uniaxial magnetic ansisotropy $K_U$ as a function of the ratio of the lattice parameters for all films. Its maximum is at around $c/a/\sqrt{2}=1.28$, which is the value for the thicker, unstrained films.}
\label{fig:K_U}
\end{figure}

\textit{Out of plane} (OOP) and \textit{inplane} (IP) magnetic measurements have been conducted with a \textit{MPMS SQUID-VSM} where H$_{max}$=7T. All OOP scans saturate at higher fields around H$=$30kOe. The IP curves mainly don't saturate distinctively, which leads to a problem of a correct subtraction of the substrate background signal. With films $>$30nm it is convenient to measure a piece of reference substrate IP, weigh it and estimate the mass susceptibility, which can then be used to calculate the substrate contribution of the sample by weighing the sample and neglecting the weight of the film. However, when measuring ultrathin films ($<$10nm) the contribution of the specimen to the overall measured signal can be negligibly small compared to the diamagnetic MgO substrate. The induced current in the pickup-coils of the SQUID is sensible to the precise substrate position, size and shape\cite{Coey06}. Therefore a different position of a sample creates a different background signal in any repeated scan. To give an example, consider an IP scan and some linear reference curve used for correction: if the diamagnetic slope of this reference curve varies around 2$\%$ due to positioning, a simple linear reference background subtraction will lead to an error in M$_{sat}$ of $\approx$16$\mu$emu (when using a 4.2x4.2mm$^2$ MgO piece), which is about the same magnitude as the signal of a 3nm thin film when using a similar sized sample piece. The less signal from that actual sample, the larger is the error induced by this method, which makes it impractical for our purposes.\\
To overcome this problem effectually, all IP scans are expected to saturate when H$>$65kOe. It has been reported, that thicker films (30nm) Mn$_{2.5}$Ga are hard to saturate\cite{Mizukami09}, so this method will tend to even underestimate H$_a$ for those films. However, we consider this a reasonable approach, since the anisotropy field for Mn$_{2}$Ga is reported to be H$_a\approx$8T for a 66nm thick film\cite{Coey11_02}. Wu et al. report H$_a\approx$6T for 30nm Mn$_{65}$Ga$_{35}$ films on Cr or Cr/Pt buffer\cite{Wu11}, which is the same compound as in our case (Mn$_{1.86}$Ga). When neglecting the Co doping, Mn$_{1.9}$Co$_{0.1}$Ga is somewhere between the two cited compositions, wherefore it should be that 6T$<$H$_a$$<$8T. A linear background subtraction was then performed around the maximum field of 7T. The estimation of H$_a$ now mainly depends both on the linearity of the IP curve in higher fields and the IP curve shape itself. Since the estimation of H$_a$ is therefore rather crude a large error was included. Furthermore the geometry of the pick-up coils in the SQUID-VSM and the differing demagnetization fields for the OOP and IP alignment will reduce the signal of any IP scan to around 80$\%$ of the same sample scanned OOP\cite{Coey06}. Therefore the IP magnetic curves had to be scaled, so that the IP and OOP scans show the same saturation magnetization M$_s$.\\
Selected hysteresis curves are shown in fig.\ref{fig:M_Cr1_Pt} for Mn$_{1.9}$Co$_{0.1}$Ga on Cr and Cr/Pt buffer. Fig.\ref{fig:M_Cr2} shows M curves for Mn$_{1.86}$Ga on Cr. All 3nm thick films show a dominating istropic contribution when measured OOP, indicating a loss of PMA. Some of the Cr buffered films showed a superimposed moment with almost zero hysteresis and small saturation field of $\approx$1kOe in both IP and OOP measurements.
This isotropic moment systematically vanishes over time for the Cr buffered films, whereas it stays unchanged for the Cr/Pt buffered ones (not shown here). A canted magnetic moment was also previsously reported for Mn$_{3}$Ga\cite{Coey11_01} and for Mn$_{2.1}$Ga\cite{Li13}. The effect here is probably related to interface effects, oxidation or paramagnetic islands that form in the initial growth of the film.
\\
M versus film thickness is shown in fig.\ref{fig:K_U}. For 30nm Mn$_{1.86}$Ga it is a factor 2 higher as previously reported\cite{Mizukami12}.\\It is known that disorder in Mn$_{3-x}$Ga thin films affects the total magnetic moment\cite{Glas13}. M appears to be constant even within the strained range of thickness, therefore no significant site disorder is occuring with the decreasing film thickness. 
However, M drops to zero for Mn$_{1.86}$Ga or reduces to the value $100\,emu/cm^3$ for Mn$_{1.9}$Co$_{0.1}$Ga as soon as the maximal observed strain is reached. When comparing the Cr buffered Mn$_{1.9}$Co$_{0.1}$Ga with the Cr/Pt buffered films the latter however don't show any significant reduction of M when going below 10nm, since they are not strained. The Cr/Pt buffered samples altogether show a higher M.\\
Following the idea of the Co doping mentioned above one notices that M for Mn$_{1.86}$Ga is roughly doubled compared to Mn$_{1.9}$Co$_{0.1}$Ga (both on Cr buffer). It is highly doubted that this effect is related to the 0.04 difference in Mn, a more realistic change related to this would be around 5$\%$\cite{Mizukami12}. So this effect has its origin in the added Co, thus reducing M$_s$ as it had been previously suggested\cite{Alijani11,Winterlik12,Ouardi12}. However, the reduction observed here is more dramatic and arguing with the absence of the (002) reflex for Mn$_{1.9}$Co$_{0.1}$Ga on Cr buffer we believe this is due to different crystallinity through different composition-dependent optimal deposition temperatures.

The uniaxial magnetic anisotropy was calculated via K$_U$=H$_a\cdot$M$_{s}$/2+2$\pi$M$_s^2$ (in cgs units; the last term accounts for the shape anisotropy in order to extract the effective magnetocrystalline anisotropy) and is plotted in fig.\ref{fig:K_U}. Although unstrained, the Cr/Pt buffered films even so show a reduction of PMA when decreasing the film thickness. However, in comparison with the strained Cr buffered films the PMA loss is less dramatic: 6nm (on Pt) Mn$_{1.9}$Co$_{0.1}$Ga still show M$_{sat}=$300emu/cm$^{3}$ and H$_a\approx$50kOe while 6nm (on Cr) has M$_{sat}=$120emu/cm$^{3}$ and H$_a\approx$5kOe. The Mn$_{1.86}$Ga films on Cr buffer however show a linear reduction of K$_U$ with decreasing $c/a$ ratio. Since the change in $c$ is dominating over the change in $a$, the loss of anisotropy can directly be related to the reduction of the unit cell height.


In conclusion it has been shown, that strain in Mn-Ga thin films reduces the magnetic anisotropy. The function of strain versus film thickness appears to be sensitively dependent on composition, i.e. Co concentration: the region of lattice distortion in Mn$_{1.86}$Ga ultrathin films on Cr buffer is almost 5nm shifted towards smaller thicknesses, when compared with Mn$_{1.9}$Co$_{0.1}$Ga on Cr buffer. The strain is reduced by increasing the Mn-Ga film thickness or by introducing a Pt buffer layer. Therefore the integration of Mn-Ga thin films into devices which demands a film thickness of t$<$10nm is only possible with the appropriate buffer layer, e.g. Pt. Conclusions drawn from thicker Mn-Ga films (30nm) should be regarded with suspicion when discussing about actual implementation into devices when t$<$10nm.\\

\begin{acknowledgments}

We would like to thank Andrew Kellock for the RBS/PIXE analysis.
Financial support by the DFG-JST (project P~1.3-A in research unit FOR 1464 {\it 
ASPIMATT}) and by the ERC Advanced Grant (291472 {\scshape Idea Heusler}) is gratefully acknowledged.

\end{acknowledgments}
%

\end{document}